\documentclass{ws-procs9x6}

\begin{document}

\title{ LARGE-SCALE RADIO STRUCTURE IN THE UNIVERSE: 
GIANT RADIO GALAXIES }

\author{M. Jamrozy and U. Klein}

\address{Radioastronomisches Institut der Universit\"{a}t Bonn, \\
Auf dem H{\"u}gel 71, D-53121 Bonn, Germany\\  
E-mail: mjamrozy@astro.uni-bonn.de}

\author{J. Machalski}

\address{Obserwatorium Astronomiczne Uniwersytetu Jagello\'{n}skiego, \\ 
ul. Orla 171, Pl-30244 Krak{\'o}w, Poland }  

\author{K.-H. Mack}

\address{Istituto di Radioastronomia, \\ 
Via P. Gobetti 101, I-40129 Bologna, Italy}

\maketitle

\abstracts{Giant radio galaxies (GRGs), with linear sizes larger than 1~Mpc 
(H$\rm_{0}=50$~km~s$^{-1}$~Mpc$^{-1}$), represent the biggest single objects in 
the Universe. GRGs are rare among the entire population of radio galaxies (RGs)
and their physical evolution is not well understood though for
many years they have been of special interest for several reasons. 
The lobes of radio sources can compress cold gas clumps and trigger star or even dwarf
galaxy formation, they can also transport gas from a host galaxy to large distances and
seed the IGM with magnetic fields. Since GRGs have about 10 to 100 times larger sizes than 
normal RGs, their influence on the ambient medium is correspondingly wider and is pronounced on 
scales comparable to those of clusters of galaxies or larger. Therefore {\it giants} could play 
an important r\^{o}le in the process of large-scale structure formation in the Universe.\\
Recently, thanks to the new all sky radio surveys,  significant progress in searching for 
new GRGs has been made.}

\vspace{-6mm}
\noindent
The main goal of our present research is to build a homogeneous data base of
all known GRGs and to use the information in a comprehensive study 
of various physical properties of the entire population.
Our sample of GRGs comes from a compilation performed 
by Ishwara-Chandra \& Saikia [1] as well as from three recent systematics 
searches for new {\it giants} [2, 3, 4]. They are supplemented with a few additional 
sources taken from other papers. The list of the GRGs will be published later\footnote
{The list of GRGs one can obtain from: {\sf  http://www.astro.uni-bonn.de/$\sim$mjamrozy}}. 
Some preliminary  results from statistical studies of the properties of this GRG sample 
are the following:
\begin{itemize}
\item{The sample contains 125 sources, 106 of which have FRII-type structures, 
8 are of FRI class and the remaining 11 have a hybrid FRII/FRI morphology.}
\item{Only 11 giants of the entire sample are identified with quasars.}
\item{Surprisingly, about 66\% of the GRGs reported in [2] and [3] do not show 
any emission lines in their optical spectra and can be classified as low-excitation radio 
galaxies (LERGs).} 
\item{The bulk of GRGs are external of dense clusters.} 
\item{7 GRGs are of a double-double type. Objects of this class show two 
pairs of lobes, presumably originating from an old and a new cycle of activity [5].}
\item{Only about 18\% of the presently known GRGs have negative declinations, 
and the majority of them are nearby objects with high flux density. Therefore, one can expect 
a large number of yet undetected GRGs in the southern hemisphere.}
\end{itemize}
\vspace*{-13pt}
\begin{table}[h]
\tbl{Mean values of physical parameters of GRGs}
{\footnotesize
\begin{tabular}{@{}lr@{}}
\hline
synchrotron age t                                    & $10^{8}$~[yr]\\[1ex]
entire volume V                                      & $10^{65}$~[m$^{3}$]\\[1ex]
equipartition magnetic field B$\rm_{eq}$             & 0.3~[nT]\\[1ex]
cocoon pressure  p$\rm_{c}$                          & $10^{-13}$~[N~m$^{-2}$]\\[1ex]
jet power Q$_{0}$                                    & 7$\times10^{38}$~[J~s$^{-1}$]\\[1ex] 
total energy delivered by the twin jets E$\rm_{tot}$ & 4$\times10^{54}$~[J]\\[1ex]
energy stored in the cocoon U$\rm_{eq}$              & 3$\times10^{53}$~[J]\\
\hline
\end{tabular}\label{table1}}
\vspace*{-13pt}
\end{table}

\noindent
Table~1 shows mean values of some physical parameters of GRGs which are derived 
from the observational data and the models of their dynamical evolution [6].

\end{document}